\documentclass[preprint,12pt]{elsarticle}

\usepackage{amssymb}
\usepackage{amsmath}
\usepackage{multirow}
\usepackage[dvipsnames]{xcolor}
\usepackage[dvipsnames]{xcolor}
\usepackage{graphicx}
\usepackage{float}

\usepackage{booktabs}

\journal{Information and Software Technology}

\begin{document}

\begin{frontmatter}



\title{StartFlow: From Method Conception to Multi-Perspective Evaluation in UX Prototyping for Software Startups}

 \author[label1]{Guilherme Corredato Guerino}
 \author[label1]{João Pedro de Souza Olivo Tardivo}
 \author[label2]{Renato Balancieri}
 \author[label2]{Gislaine Camila Lapasini Leal}
 \affiliation[label1]{organization={State University of Paraná},
             city={Apucarana},
             state={Paraná},
             country={Brazil}}
 \affiliation[label2]{organization={State University of Maringá},
             city={Maringá},
            state={Paraná},
             country={Brazil}}

\begin{abstract}
\textbf{Context.} Software startups face significant challenges in building minimum viable products, particularly in the early stages, when resources are limited and expertise in user experience is scarce. \textbf{Objective.} Introduce StartFlow, a structured method that helps non-specialized professionals create MVP prototypes using the wireflow technique, a combination of wireframes and user flows. StartFlow consists of three steps: (i) organizing features; (ii) building wireflows; and (iii) verifying and refining them based on usability heuristics. \textbf{Method.} To assess the method Startflow, we first conducted a focus group with researchers in Software Engineering, Human-Computer Interaction, and Software Startups. Afterward, we conducted a proof-of-concept study, which consisted of an experiment and a heuristic evaluation with experts. \textbf{Results.} The qualitative analysis of the focus group revealed that participants found the method straightforward, flexible, and helpful in structuring user flows and identifying visual components. However, they also pointed out the need to improve its presentation, clarify its iterative nature, and strengthen its connection to broader UX principles. The results of the proof-of-concept indicate that participants who used StartFlow created clearer prototypes, adhered to the proposed user stories and business rules, and presented fewer usability defects. Furthermore, the method was well evaluated for its ease of use and intended future adoption. \textbf{Conclusion.} The study reinforces the potential of StartFlow as an accessible tool to support user-centered development in software startups from the earliest stages of their product development.
\end{abstract}

\begin{graphicalabstract}

\end{graphicalabstract}

\begin{highlights}
\item The StartFlow method supports UX in startups with limited expertise in the field;
\item StartFlow combines wireframes and user flows in a structured approach;
\item Experiment shows greater clarity and adherence to requirements with the use of StartFlow;
\item Heuristic evaluation indicates fewer defects in prototypes created with StartFlow;
\item Participants reported greater intention to use and satisfaction with the proposed method.
\end{highlights}


\end{frontmatter}

\section{Introduction}\label{introduction}

Software startups emerged as a response to the market's need for innovative and efficient solutions, operating in an environment characterized by volatility and the search for sustainability \citep{giardino2014we,paternoster2014software}. These companies face unique challenges, operating with lean teams and limited resources while navigating a dynamic and technologically evolved market \citep{giardino2015key,paternoster2014software,rafiq2021analytics}. The lack of operational history, the scarcity of resources, and the constant evolution of technologies characterize this challenging scenario, where software startups seek to solve existing problems and stand out in a competitive market.

In the early stages of their existence, software startups face several challenges contributing to high failure rates \citep{giardino2015key, leal2020, guerino2021user}. Inconsistency in understanding market problems and proposing solutions, the difficulty in obtaining capital, and the lack of demand for the product are crucial factors contributing to these failures. In this context, the need to create viable, sustainable, innovative business models becomes even more pressing.

To deal with challenges in the early stages, the Lean Startup methodology \citep{reis2011lean} stands out as an approach that emphasizes validating ideas through the Minimum Viable Product (MVP). MVP allows software startups to learn from the market in an agile and efficient way, quickly adapting to user needs and feedback. This iterative, customer-centric approach allows startups to test their market hypotheses pragmatically, minimizing risk and maximizing learning. Thus, the literature investigates aspects of the construction of MVPs in software startups, for example, which definitions are being used \citep{lenarduzzi2016mvp}, what they understand as a minimum \citep{shepherd2021lean}, and how Software Engineering (ES) has supported this construction \citep{alonso2021systematic}. Therefore, some proposed themes can help software startups evolve, such as technical debt management, risk management, and user experience (UX) \citep{unterkalmsteiner2023software}. In this sense, studies state that UX in software startups, despite having a challenging application, promotes a competitive advantage for these companies, as they focus the development process on a user-centered approach \citep{sofia_uxleris:2021,guerino:2022, choma:2023,guerino2023}. UX is defined as the user perceptions and responses resulting from using a system, product, or service \citep{iso:2019}, highlighting the importance of putting the user at the center of design and development decisions. Software startups face challenges in effectively applying UX practices despite the potential benefits, mainly due to a lack of resources and UX expertise within teams \citep{hokkanen2015ux,saad:2021,CHOMA2022107041}.

As far as we know, few studies provide specific guidance for applying UX in these companies, especially considering resource limitations and the lack of UX specialization in teams. Hokkanen et al. \citep{hokkanen2016minimum} provide a framework that focuses on the pillars of attractiveness, accessibility, professionalism, and sales of a product idea. The Minimum Viable User Experience (MVUX) framework was built based on Finnish culture and aims to collect feedback for the initial product design focused on UX aspects. Another study that addresses UX practices in the context of software startups proposes Lean UX, a development process that seeks to encourage the construction of products with a focus on UX \citep{gothelf2013lean}. This process combines design thinking, Lean Startup, and agile methodologies, focusing on prototyping and rapid product validation. However, none of these proposals define UX guidelines for software startup professionals who want to focus on a specific startup activity, such as building an MVP. Furthermore, specialized collaborators are needed to work with the practices or sufficient resources to outsource the work, which is difficult in startups in the early stages. In the work of \cite{lermen2023does}, the authors state that the startups investigated that used Lean UX were due to their higher levels of maturity. Then, while approaches like Lean UX and MVUX offer important strategic principles, they do not provide operational and phase-specific steps to help professionals translate user stories or initial ideas into structured flows, prototypes, or MVPs. StartFlow seeks to fill this gap by offering specific guidance for this activity.

This knowledge gap represents a significant research opportunity, seeking to develop innovative methods and processes that support software startup professionals in the efficient and results-oriented application of UX practices, contributing to a better user experience from the first versions of the product and increasing these companies' chances of survival in the competitive software startup market.

Therefore, the goal of this article is to introduce Startflow, a method that aims to help startup professionals create a prototype of their MVPs based on a UX technique called wireflows (a combination of wireframes and user flow) \cite{guerino2024wireflows}. Startflow was designed based on previous studies in the literature and also based on studies that explored the practice in the software startup industry \cite{guerino2021user, guerino:2022, guerino2023, guerino2024wireflows, guerino2024investigating, guerino2025towards}. Startflow consists of three steps: i) understand and organize the MVP's functionalities; ii) build the wireflows; iii) verify and refine the created wireflows. The method also provides the objective, input, description, question points, and expected output for each of its steps, so that the professional ultimately has a wireflow for their MVP.

In addition to introducing Startflow to the academic community, this study also presents the results of a focus group discussion with HCI and Software Engineering researchers on the method, as well as a proof of concept we conducted using the method based on an experiment and a heuristic evaluation. First, in the experiment, we simulated two software startups and asked participants to create prototypes for an application. The experimental group used Startflow, and the control group used only paper prototyping. At the end of this experiment, we analyzed the components used in each prototype and collected participant perceptions using the Technology Acceptance Model (TAM) \cite{venkatesh2000model}. Second, we invited nine experts in Human-Computer Interaction (HCI) and Software Engineering (SE) to conduct a heuristic evaluation-based inspection \cite{nielsen1994usability} of the prototypes created in the experiment by both groups. Then, we compared the inspection findings for the control and experimental groups.

\section{Related work}\label{relatedwork}

Research involving UX practices in software startups is recent and has been growing in recent years. Hokkanen and  Vaananen-Vainio-Mattila \cite{hokkanen2015ux} interviewed eight startups in Finland to understand their practices and future UX work needs. Based on the results, the authors highlight some contributions for startup professionals, such as the need to be skilled in collecting and analyzing user information, applying quick interviews, research and testing methods, searching for users, preparing for feedback and data they will receive, and create a UX strategy. Guerino et al. \cite{guerino2021user} also interviewed six professionals from early-stage software startups to explore how UX work was carried out in these companies. The authors identified that startups have specific reasons for using the practices, such as supporting the construction of the MVP and promoting mentoring programs.

Nguyen-Duc et al. \cite{nguyen2017influences} conducted a multiple case study with twenty European startups to investigate which factors influence prototyping work in these companies. The main result of this study indicates that the artifact adoption approach influences disposable and evolutionary prototypes, the team's competence, and collaboration and involvement with the customer. The authors conclude the article by mentioning that future work can investigate startups' different learnings during the prototyping process and the importance of this learning for the product created.

Silveira et al. \cite{sofia_uxleris:2021} surveyed 88 startup professionals to identify how startups deal with UX work and how relevant UX is to these professionals. As a result, the authors indicate some challenges in UX work for startups, such as combining UX work with agile practices, making practices leaner for UX work, and training and developing skills to carry out UX activities. In the same sense, Guerino et al. \cite{guerino2023} surveyed 90 startup professionals to verify the perception of usefulness that these professionals have about UX techniques and methods. The authors found that interviews are the most used method, but usability testing, competitive analysis, and high-fidelity prototyping are considered most useful for startups at different stages of maturity.

The studies investigate UX practices in software startups and show that, as in any software company, practices can be a decisive factor in the success of products. Studies involving the concepts of UX and startups are recent and investigative, generally exploring factors and points of difficulty for these companies, stating that more research is needed to contribute to the state of the art. However, to the best of our knowledge, the literature lacks studies with specific methods for startups that encourage the application of some UX techniques. Then, this paper contributes to the state of the art by describing Startflow, which encourages using the UX wireflow technique, combining wireframe and user flow.

\section{StartFlow method description}\label{starflow}

To design and achieve this version of Startflow, several studies were necessary to map the needs of the software startup industry and also to understand the state of the art. To this end, we conducted a systematic literature review, which provided guidance on the UX methods and techniques used in software startups based on academic reports \cite{guerino:2022}; a survey of software startups, which allowed us to identify the practice of UX work in these companies \cite{guerino2023}, as well as identify their main challenges \cite{guerino2024investigating} and the relationship between the professional's experience and this work \cite{guerino2025towards}; a study based on a sequence of interviews with several startup managers and directors, to understand these professionals' perceptions of UX work \cite{guerino2021user}; and finally, an initial experiment with the wireflow technique in a real case study at a software startup \cite{guerino2024wireflows}. All of these studies were crucial for us to design a method that had both an academic and well-founded foundation, as well as practical relevance in the context of software startups. The survey findings on UX challenges in startups helped define the method's focus on simplicity, low cost, and minimal time investment. The insights into professionals' perceptions of UX work obtained from the interview reinforced the need for practical question points to guide decision-making. Finally, the results of our wireflow experiment in a real startup inspired the method's emphasis on structuring initial functionalities through visual flows, rather than isolated screens.

Unlike Lean UX \citep{gothelf2013lean} and MVUX \cite{hokkanen2016minimum}, which offer broad processes and rely on cross-functional collaboration, StartFlow focuses specifically on translating early user stories into low-fidelity prototypes. Startflow is not intended to replace iterative UX cycles, but rather to operationalize a three-step method designed for teams with little or no UX experience who need to organize their initial resources and produce coherent early prototypes quickly. Thus, StartFlow complements existing UX processes focused on startups.

\begin{figure}[ht]
    \centering
    \fbox{\includegraphics[width=0.9\textwidth]{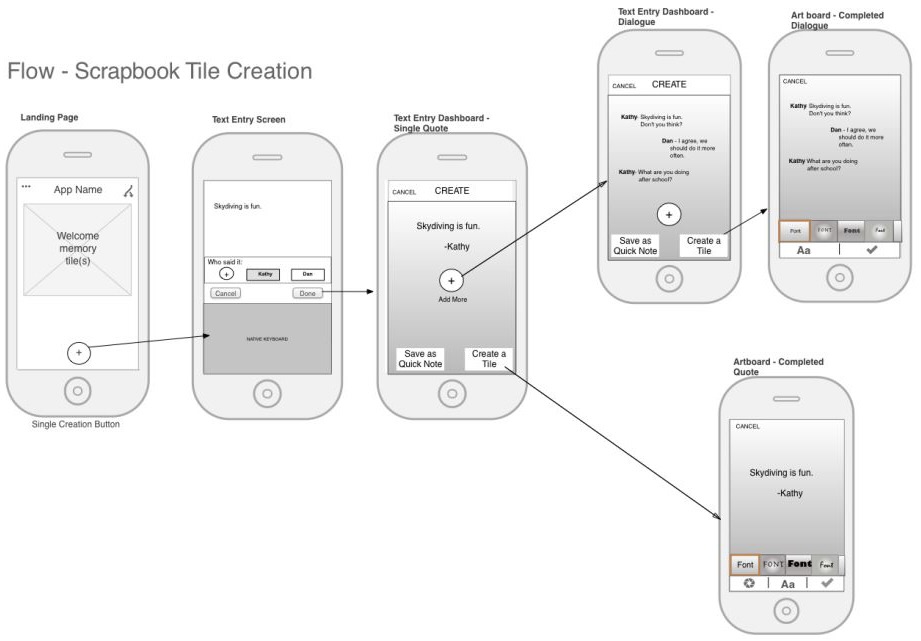}}
    \caption{Example of wireflow created to represent scrapbook tile creation \cite{laubheimer2016}.}
    \label{fig:wireframesref}
\end{figure}

Therefore, StartFlow is a method that aims to support software startup professionals in building MVPs using wireflows, a UX method resulting from the combination of wireframes (low-fidelity prototypes) and user flow maps. Wireflows are artifacts that combine low-fidelity wireframes with user flow diagrams, representing both the visual structure of screens and the transitions between them. They aid in initial reasoning about navigation logic and task completion. Figure \ref{fig:wireframesref} illustrates a simplified wireflow containing screens, interaction points, and directional transitions that guide the user through tasks. Note that, although Figure \ref{fig:wireframesref} shows a wireflow created in a digital tool, this artifact can also be constructed with paper and pen.

With StartFlow, we seek to help software startup professionals deliver a better UX in the early versions of their products, in an agile, low-cost way. Furthermore, professionals will obtain a first version of the visual representation of the product already verified and refined based on the UX points inserted in the method. StartFlow can be used when the software startup has already gone through an idea or business model validation process and wants to explore how the features designed for the product will be designed.

To achieve this, StartFlow can be used by just one person assigned to the job or, if it is compatible with the startup's structure, by a group working together to apply the method. The professional who will use the method can choose specific features to use in the method. This way, if all the product's features have not yet been defined, the method can be used for those already defined. As the professional identifies more features for the product, the method can be replicated at other times.

To make this possible, StartFlow consists of three steps: (i) understanding and organizing the features; (ii) building the wireflows; and (iii) verifying and refining the wireflows. Each step consists of its objective, the necessary input, the description, the questioning points, and the expected output. 
In the following subsections, each of its steps is detailed.

\subsection{Step 1. Understanding and Organizing the Features}

\textbf{Objective.} Organize the features that will have their visual representation created by StartFlow. If there are no defined features, this step also helps to identify them.

\textbf{Input.} Artifact (formal or informal) that represents the features, such as notes created digitally or through post-it notes, records on whiteboards or Kanban, documentation organized in backlog tools, templates, or formal requirements documents.  

\textbf{Description.} Before building the wireflows, it is necessary to organize the features that will have their visual representation created by StartFlow. The structure in which the requirement or user story is represented, whether formal or informal, low or high level, does not interfere with using StartFlow, considering that the professional knows what feature is represented. Therefore, the professional must use the questioning points in this stage of StartFlow to organize the features in order of priority. If the professional does not have any formally or informally documented artifact that represents the product's features, he or she must identify the features used in constructing the wireflows. To simplify this step, high-level user stories are recommended to be documented informally on a whiteboard, post-it notes, or digital documents. To document these features as they emerge, even informally, the Connextra template \citep{cohn2004user}, the most commonly used template for documenting user stories \citep{lucassen2016use}, is recommended.

\textbf{Questioning points.} The following questions can be used to select and organize features or identify new features:

\begin{itemize}
    \item "Have we selected the features we want to create in the wireflow? Is there any other feature that could be used at this time?"
    \item "Of the selected features, which is the most important? And, the next one?" - Carry out this question until all features are organized by priority.
    \item "In the first contact with the application, what tasks will the user be able to perform?"
    \item "For users who already have experience with the application, what tasks will they be able to perform?"
    \item "What market demand will the application meet? In what way?"
    \item "Are there competitors in the market? What tasks will be similar? What will be innovative about the application?"
\end{itemize}

\textbf{Output.} At the end of this stage, an artifact (user stories documentation) will be produced that includes the features, organized by priority, to be used in the following stages of StartFlow.

\subsection{Step 2. Building the Wireflows}

\textbf{Objective.} Build wireflows visually representing the features identified and organized in Step 1. This separation by feature will help the professional have the user's vision for each of the possible operations in the application and verify improvements to the wireflows for each feature.

\textbf{Input.} Artifact with the features generated in Step 1.

\textbf{Description.} With the artifact containing the features represented and organized by priority, the professional will build the wireflows that visually represent each. If the professional knows digital prototyping tools, these can be used. However, if the professional does not have experience, he or she can build the wireflows with paper and pen. To help with this stage of StartFlow, categories of elements that can be used during the construction of wireflows were created, which are \textit{layout}, \textit{triggers} and \textit{connectors}. These categories of elements are detailed in Table \ref{tab:elements}. To aid creativity and the use of category elements, professionals can explore widely used user interface pattern libraries (e.g., Material Design\footnote{https://m3.material.io}, Apple Developer\footnote{https://developer.apple.com/design/resources/}) to represent standard interface components consistently during wireflow construction. These libraries are not mandatory, serving only as optional references to facilitate the sketching process.

\begin{table}[ht]
    \footnotesize
    \centering
    \begin{tabular}{cp{8.5cm}p{2cm}}\hline
     \textbf{Element} & \textbf{Description} & \textbf{References} \\\hline
      Layout & Elements that are part of the organization of the screen as a whole, for example, information containers, image collections, sliders, lists and tables, text/data insertion fields, cards, and checkboxes, among others. & Material Design and Apple Developer \\\hline
      Triggers & Elements that trigger a call to another application screen, for example, buttons with texts, buttons with icons, and hyperlinks. From these components, the application flow will be represented. & Material Design and Apple Developer \\\hline
      Connectors & Elements used to join the screens that are part of the flow triggered by the trigger, represented by an arrow. The connector must have a trigger as the exit point and a screen as the arrival point. & \citep{aleryani2016comparative,wulandari2017design} \\\hline
    \end{tabular}
    \caption{Category of elements used to create wireflows.}
    \label{tab:elements}
\end{table}

\textbf{Questioning Points.} The following questions can be asked to help the professional create wireflows:

\begin{itemize}
    \item “How many screens does it take for the user to execute this feature from start to finish?”
    \item "What are the necessary elements on each screen for the user to perform this feature?”
    \item “Which other screen will the trigger on one screen take the user to?”
    \item “If the user enters information incorrectly, what will happen to the flow of feature in the application?”
    \item If the user wants to return to a previous screen, is it possible?”
    \item “Are there ways for the user to perform this feature with fewer clicks?”
    \item “What will happen to the application flow when the user completes the feature?”
\end{itemize}

\textbf{Output.} A set of wireflows containing layout elements, triggers and connectors that visually represent each feature.

\subsection{Step 3. Verifying and Refining the Wireflows}

\textbf{Objective.} Check and refine, if necessary, each of the wireflows. This way, the professional will be able to reflect on the user's flow within the system. Furthermore, verification will identify and correct flow and UX issues, promoting the evolution of wireflow and driving better interaction design.

\textbf{Input.} Wireflows created in Step 2.

\textbf{Description.} After building the wireflow for a specific feature, the professional must check whether the created wireflow needs improvement and refinement before building the wireflow for the following feature. To do this, the professional must consult the questioning points in this step to verify the wireflow created. Furthermore, explanations are proposed for each questioning point to assist the professional. If the professional answers any of the questions negatively, the wireflow must be refined, and improvements must be proposed until a positive answer is reached for all questions. The questioning points and their explanations were created following Nielsen's Heuristics \citep{nielsen1994usability} and aimed to ensure that the visual representation already projects a flow of information that promotes a better user experience.

\textbf{Questioning Points.} Table \ref{tab:heuristics} shows the heuristics explored, the question, and the explanation of each questioning point in this step.

\begin{table}[htp]
    \footnotesize
    \centering
    \begin{tabular}{p{2.5cm}p{4cm}p{6cm}}\hline
     \textbf{Heuristic} & \textbf{Question} & \textbf{Description} \\\hline
    Flexibility and efficiency of use & Do all screens have at least one trigger for the user to activate? & There must be at least one trigger (even if it is to return to the previous screen) so the user is not locked on the same screen forever\\\hline
    Visibility of system status & Towards the end of the feature, is there a screen that provides feedback indicating the completion of the task? & There must be an indication to the user that the task they were performing has finished, providing clear feedback\\\hline
    Match between system and the real world & Are triggers that have texts adequately described? & Buttons with texts or hyperlinks present the correct text description, allowing the user to understand the action clicking the trigger will perform.\\\hline
    Match between system and the real world & Are icon/image-based triggers adequately represented? & Buttons with icons or images correctly display the icon that allows the user to understand the action clicking the trigger will perform\\\hline
    Error prevention & If there are fields for data entry, are the mandatory fields marked? & It is essential to define which fields are mandatory in possible product screens\\\hline
    Help users recognize, diagnose, and recover from errors & Does the interaction flow consider errors that users may make? & Error screens must be created, and interaction with them must be well-defined. If necessary, an alternative flow must be created containing error screens based on triggers that activate these screens\\\hline
    Flexibility and efficiency of use & Are there triggers for the user to undo an action or return to the previous screen? & It is essential that the user has the control and freedom necessary to redo or undo actions and correct possible mistakes they may have made\\\hline
    Consistency and standards & Are all connectors starting from triggers? & Only triggers can trigger a call to another screen, including alternative flow screens for errors\\\hline
    \end{tabular} 
    \caption{Heuristics, questions and description of each questioning point in Step 3 of StartFlow.}
    \label{tab:heuristics}
\end{table}

\textbf{Output.} At the end of the construction, verification and refinement of all feature wireflows, the professional will have an initial set of visual representations of their MVP, with the necessary screens to guarantee the flow designed for the application and initial validation of the user stories, defined requirements, or features.

\section{Focus Group}

The focus group is a qualitative technique used to collect data through a group interview. The steps used in this study followed the recommendations of other focus groups carried out \citep{Kontio2008, deFranca2015}, and were organized into: (i) research definition; (ii) focus group planning; (iii) execution of the focus group; and (iv) data analysis and reporting of results.


\subsection{Research definition}

The focus group is one of the most suitable techniques to obtain initial feedback on new concepts, how models or proposals are documented, and discover underlying motivations about the object of study \citep{Edmunds2000, Kontio2008}. In this sense, the object of study of this focus group is the StartFlow method. The goal of the study was defined according to the \textit{Goal-Question-Metric} \citep{Basili1994} paradigm and is shown in Table \ref{tab:GQM}.

\begin{table}[ht]
    \footnotesize
    \centering
    \begin{tabular}{p{3.8cm}p{9cm}}\hline
    \textbf{Analyze} & the StartFlow method\\
    \textbf{for the purpose of} & characterize\\
    \textbf{with respect to} & its applicability in creating wireflows in software startups\\
    \textbf{from the viewpoint of} & researchers in Software Engineering, HCI and Software Startups\\
    \textbf{in the context of} & academic\\\hline
    \end{tabular}
    \caption{Focus group goal according to the \textit{Goal-Question-Metric} paradigm.}
    \label{tab:GQM}
\end{table}

To achieve this goal, we proposed a research question (RQ) that guided the entire process of planning, executing, and analyzing the results of the investigation: \textit{"What is the participants' perception about the usefulness, ease of use, and understanding of the StartFlow method?"}

\subsection{Planning}

\textbf{\textit{Participants selection.}} The size of a focus group can vary from 3 to 12 participants but typically has between 4 and 8 participants \cite{Kontio2008}. For this focus group, we selected seven researchers in Software Engineering, HCI or Software Startups as participants. Participants were selected using convenience sampling, who accepted the invitation according to their availability to participate in the study. We contacted participants via \textit{e-mail}, and the common date was obtained using the Doodle\footnote{https://doodle.com/en/} platform. In addition to the participating researchers, the study also had the participation of the first author as moderator of the focus group session.

\textbf{\textit{Segmentation.}} We divided the participants into two groups, \textit{lovers} and \textit{haters}. The lovers' group comprised three participants, while the haters' group comprised the other four participants. This division was carried out according to the participants' experience, aiming to balance the groups.

\textbf{\textit{Design and Strategy.}} The design used for this focus group was based on the dynamic ``\textit{lovers} and \textit{haters}'' \citep{Colucci2007}. In this dynamic, participants are divided into two groups with predefined roles in the study: the \textit{lovers} group will argue in favor of the StartFlow method, while the \textit{haters} group will argue against it. This type of dynamic aims to generate a discussion between the groups who will oppose their opinions, rendering the data that will be analyzed later. Thus, the strategy that was followed during the study was organized in the following steps:

\begin{enumerate}
     \item Separation of groups and explanation of the role of each one of them;
     \item Explanation of StartFlow, its steps, composition, and purpose;
     \item Internal discussion of the subgroups about the positive (lovers) and negative (haters) points. In this step, each subgroup must prepare post-its that they will place on a board;
     \item The participants place all the post-its on the board and comment on each one;
     \item After inserting all the post-its on the board, participants should comment on each one. The moderator should promote a debate between lovers and haters as post-its are read.
\end{enumerate}

\subsection{Execution}

The participants and the focus group moderator met virtually in a \textit{Google Meet} room. In the first moment of the meeting, the moderator asked all participants to read and accept, if they agreed, the Informed Consent Form (ICF) and fill out the characterization questionnaire. The characterization questionnaire had questions about the participants' background, age group, gender, and experience with software startups, Software Engineering, and User Experience. Table \ref{tab:demographicsFG} shows the participant's demographic information.

\begin{table}[ht]
    \centering
    \footnotesize
    \begin{tabular}{p{12.5cm}r}\toprule
    \textbf{Characteristics} & \textbf{N}\\\midrule
    \textbf{Educational background} & \\
    Ph.D. student & 4\\
    Ph.D. & 2\\
    M.Sc. & 1\\
    \textbf{Age range} & \\
    From 26 to 35 & 4\\
    From 20 to 25 & 2\\
    Over 45 & 1\\
    \textbf{Gender} & \\
    Female & 5\\
    Male & 2\\
    \textbf{Experience with Software Startups} & \\
    High (I am a researcher in the area of software startups and I aim to propose solutions for these companies) & 4\\
    Medium (I have good knowledge about software startups; I have done some reading about and participated in projects with startups) & 1\\
    Low (I have reasonable knowledge about software startups but have never researched or participated in any startups) & 2\\
    \textbf{Experience with User Experience (UX)} & \\
    High (I have experience in UX design or research in the software industry) & 4\\
    Medium (I have experience in UX design or research only in the classroom) & 2\\
    Low (I have UX knowledge acquired in lectures or readings) & 1\\
    \textbf{Experience with Software Engineering} & \\
    High (I have experience in software development projects in the industry) & 4\\
    Medium (I have experience in software development projects only in the classroom) & 3\\\hline
    \end{tabular}
    \caption{Study participant's demographic information.}
    \label{tab:demographicsFG}
\end{table}

According to Table \ref{tab:demographicsFG}, we noted that all participants had a master's as a minimum degree, and most were in the doctoral phase. Regarding experiences, we verified that all participants had at least average experience with some of the topics. In addition, we observed that most participants have high experience with one of the topics involved in the research.

After acceptance of the ICF and completion of the characterization questionnaire, the moderator followed the steps previously defined in the study strategy. The groups were divided into two other meeting rooms and had access to StartFlow documentation for internal discussion. To support the dynamics of \textit{lovers} and \textit{haters}, we used two virtual boards in the Miro tool\footnote{https://miro.com/pt/}, where participants could write their post-its and insert in the boards, one for the lovers' group and one for the haters' group. Each board had three columns. In the lovers' board, the columns were ``I find the StartFlow method useful because...", ``I find the StartFlow method easy to use because...", and ``The StartFlow method is easy to understand because ...". In turn, the haters' board contained the columns ``I do not find the StartFlow method useful because...", ``I do not find the StartFlow method easy to use because...", and ``The StartFlow method does not easy to understand because...".

The focus group session lasted approximately two hours. We did not record the internal discussion of the groups, respecting privacy and favoring debate among participants. However, we recorded the exposition session of the groups' opinions and the comments on the post-its so that it was possible to analyze the data obtained.

\subsection{Data Analysis}

For data analysis, we used a qualitative approach to answer the previously defined research questions. We analyzed the post-its inserted in the boards, as well as the comments made about each one of them. 
From this, the first author encodes the opinions to emerge the themes in the analysis. Subsequently, the codification and themes were revised by the second and third authors.

\subsection{Results}


From the qualitative analysis, we identified three topics that highlighted the \textit{usefulness} of StartFlow for the lovers: question marks, user flows, and visual elements. Regarding the question marks, the group mentioned that they helped them think about the features and the product requirements (see quote Q1). In addition, the method also helps to think about the user's flow during the interaction (see quote Q2). Finally, as well as helping to identify visual elements (see quote Q3), the lovers also mentioned that the method helps to reuse screen components, allowing them to be used in other projects (see quote Q4). There were no comments when the haters were asked if they disagreed with these topics.

\begin{itemize}
\small
    \item Q1: \textit{``As the product development process progresses, given the number of activities performed, some of these elements (features or requirements) may be forgotten. Therefore, the question marks help in this regard."} (P1)
    \item Q2: \textit{``By using the method, the entire interaction of the product application has a well-defined beginning, middle, and end."} (P3)
    \item Q3: \textit{``As the features are identified, we can think about the visual elements that will allow these features to be implemented in the product."} (P2)
    \item Q4: \textit{``As certain components have already been used in previous projects and are known, this can save us much time, contributing to the method's speed purposes."} (P1)
\end{itemize}

On the other hand, the haters criticized the method's usefulness according to the following topics: requirements artifact, method evaluation measures, and the connection to the user experience. Initially, the haters mentioned that the method looks fixed and sequential, which is not suitable for startups that are more agile environments (see quote Q5). At this point, there was a counterpoint from a lover participant who mentioned the possibility of going back a few steps to refine the wireflows created (see quote Q6). Following the criticism voiced by the haters, one participant mentioned that it needs to be clarified what the baseline to compare the method and how the advantages can be measured or evaluated, which casts doubt on the benefits that StartFlow can generate (see quote Q7). They also questioned the method's connection to UX since it only focuses on interface and interaction (see quotes Q8 and Q9). There was no opposition from the lovers to these last two topics.

\begin{itemize}
\small
    \item Q5: \textit{"When you look at the method, it looks like waterfall [...], which is not very much the reality of startups, something more incremental and agile. One idea would be to give it the idea of being iterative. There is a "yes" and "no" that comes back, but it does not go back to identifying the requirements."} (P4)
    \item Q6: \textit{"I do not agree that the method is similar to a waterfall; it can go back to refine the wireflows."} (P3)
    \item Q7: \textit{"Today, startups already use some things. You need to see how the metrics compare with these other artifacts."} (P4)
    \item Q8: \textit{"In the question marks, some questions could go back to the UX side. Because we are talking a lot about UI and interaction here. However, we need to focus on the experience itself."} (P5)
    \item Q9: \textit{"Even if you deal with the visual aspects, there must be a greater connection with UX. There was a lack of artifacts or guidelines similar to the question marks to make a connection."} (P7)
\end{itemize}

Regarding the \textit{ease of use} of the method, the lovers provided answers that emerged under the following topics: question marks, flexibility of the artifacts to be used, and flexibility of the prototyping tools. Concerning the question marks, the lovers mentioned that during the execution of the method, some challenges may arise and that the points can help the professional to overcome these challenges (see quote Q10). Regarding the flexibility of the artifacts, the lovers praised the fact that there is no restriction on the specific documentation of requirements or user stories to be used in the method, and formal or informal artifacts can be used in StartFlow (see quote Q11). In addition, another positive point highlighted by the lovers was about prototyping tools, since digital tools can be used if the professional already has this experience, or even pen and paper (see quote Q12). It should be noted that there was no opposition from the haters with these points raised by the lovers.

\begin{itemize}
\small
    \item Q10: \textit{"We understand that in the course of using the method, difficulties and challenges will arise, and these question marks represent an aid to these situations."} (P1)
    \item Q11: \textit{"Anything from formal documents to quick notes can be considered by the method, and we understand that this is very positive."} (P3)
    \item Q12: \textit{"The method is very flexible about prototyping tools, and this is also a positive point."} (P2)
\end{itemize}

On the other hand, the haters also raised some criticisms regarding the ease of use of the method, linked to the following topics: presentation of the method, stages of the method and requirements artifacts. Regarding the presentation of the method, the haters mentioned that it should be leaner and more dynamic in order to fit into the reality of startups (see quote Q13), in addition to the fact that the information is very textual and that other forms of visual presentation could help with use (see quote Q14). Concerning the stages of the method, the haters pointed out that the method seems systematic for the startup context and that the form of presentation should be better worked on (see quote Q15). At this point, there was a counterpoint from the lovers' group, where P1 mentioned that he realized the possibility of iteration in the method and needed to be explained better (see quote Q16). Still on ease of use, the haters pointed out the requirements artifact used in the method since it is not always possible to determine all the product's feature from the start (see quote Q17). At this point, there was a counterpoint from the lovers' group, where they mentioned that the method would be used when you already have some features defined and want to represent them visually (see quote Q18).

\begin{itemize}
\small
    \item Q13: \textit{"There is much text. I am a more visual person. Maybe bring in more images and diagrams. It would make the method even leaner."} (P6)
    \item Q14: \textit{"One thing that could help make your text leaner, one possibility you could present is to create something in Figma, or something that the person could access in a more visual way. I think it would be simpler. It would help you with how the professional will receive what you are creating."} (P7)
    \item Q15: \textit{"The way it was presented, we even see an interaction with that flowchart. However, it still seems to be systematic and sequential, and it is something that in the context of a startup, in the context of haste, and in the context of people often doing what they can at the moment, this structure can make it difficult. It is necessary to emphasize that it is agile and iterative."}(P5)
    \item Q16: \textit{"I could see that the method is iterative. However, either in the diagram or in a specific section, the method could explain the possibility of applying the method as described or making cycles within the stages optional."} (P1)
    \item Q17: \textit{"Having to talk about all the features right at the beginning is complicated because, in a startup, many things are 'thrown away'."} (P5)
    \item Q18: \textit{"The method fits when you already have the features defined or just want to represent them on the screen. It is a step after defining the features. The features are already defined; now let's define the representation."} (P2)
\end{itemize}

Concerning \textit{ease of understanding}, the lovers highlighted the clarity of the method's steps (see quotes Q19 and Q20). There was no opposition from the haters on this point.

\begin{itemize}
\small
    \item Q19: \textit{"In a short space of time, we can easily grasp the central idea of the method, which can help us adopt it daily."} (P1)
    \item Q20: \textit{"The process is not long, which makes it easier to use and understand."} (P2)
\end{itemize}

From the point of view of the haters, one topic that could make it challenging to understand the method is the word \textit{wireflow} since it is not a term widely used within the context of startups (see quote Q21). In addition, it was suggested by the haters' group that the UX vision could be inserted or better exposed within the question marks. There was no opposition from the lovers to these comments.

\begin{itemize}
\small
    \item Q21: \textit{"I have worked with non-UX people, and they talk a lot about brainstorming, design thinking, these things that everyone in a startup knows, and not necessarily such a specific practice. We are thinking of a non-UX audience."} (P5)
\end{itemize}

\subsection{Synthesis}

The main findings reveal that the group identified as ``lovers" highlighted positive aspects such as the usefulness of questioning points for reflecting on features and requirements, the clarity of the user interaction flow, and the reuse of visual components, which contributes to greater efficiency. These participants also emphasized the method's flexibility regarding the use of formal or informal artifacts and prototyping tools, as well as the clarity and objectivity of its steps, which facilitate practical adoption. On the other hand, the ``haters" group presented relevant criticisms, such as the perception that the method is excessively sequential, which is inconsistent with the iterative and agile nature of startups. They also questioned the lack of clear metrics to evaluate StartFlow's benefits, the limited connection with broader aspects of UX, and the excessive text in the method's presentation, suggesting a more visual and dynamic approach. They also mentioned that the term "wireflow" may be unfamiliar to professionals outside the UX universe, hindering its understanding and acceptance. In response to some of these criticisms, "lovers" argued that the method is iterative and more appropriate when functionalities are already defined. In summary, the study demonstrated that, although StartFlow presents qualities recognized by participants in terms of applicability and clarity, it still requires adjustments to better align with the practical and agile context of startups, particularly in its presentation and integration with broader UX principles.

\section{Proof of Concept}\label{proof}

To verify the results of StartFlow in practice, we conducted a proof of concept consisting of two studies described in this section: a experiment for creating wireflows and a heuristic evaluation of the created wireflows. The proof of concept did not involve the development of a software tool. Instead, participants applied StartFlow to create wireflows of a new application, and the resulting artifacts were subsequently evaluated through heuristic inspection.

\subsection{Experiment}\label{controlled}

The goal of the experiment was defined based on the GQM framework \cite{aversano2004framework}: ``Analyze the Startflow method for the purpose of evaluation with respect to their efficiency, perceived ease of use, perceived usefulness, and behavioral Intention to use from the point of view of non-UX professionals in the context of computer science students simulating a software startup".

\subsubsection{Planning}\label{controlled-planning}


\textbf{Context.} We conducted the experiment through group dynamics, where each group simulated being a development team of a software startup whose main product is an application for monitoring university students' complementary academic activities (CAAs). The experiment was conducted in person with 3rd and 4th-year Computer Science students at the State University of Paraná, Brazil. Each group had to prototype, through wireflows, their proposed solution for the application, following the user stories (US) and business rules (BR) of the product shown in Table \ref{tab:us_br}.

\begin{table}[ht]
\footnotesize
\centering
\begin{tabular}{cp{12cm}}
\toprule
\textbf{ID}         & \textbf{Description} \\ 
\midrule
US1           & As an undergraduate student, I want to request the use of CAAs to complete the minimum workload of 122 hours. \\
US2          & As an undergraduate student, I want to manage my CAA requests to check their status. \\
BR1        & A minimum workload of 122 hours is a requirement for graduation. \\
BR2        & CAA requests must contain supporting documents and will be approved by the course board. \\
BR3        & CAAs are divided into 13 distinct categories, each with different maximum limits for calculation (the categories and limits were provided to the subjects). \\
BR4        & Students must complete activities in at least four different CAA categories, regardless of whether they have reached the total hours stipulated in the curriculum matrix. \\
BR5        & The workload has 60 minutes as the minimum unit of time. \\
\bottomrule
\end{tabular}
\caption{User Stories and Business Rules}
\label{tab:us_br}
\end{table}




\textbf{Selection of subjects.} We selected eight computer science students to take part in the study. The participation was voluntary and occurred outside of class time. The idea was to simulate the human resources characteristics of a software startup, often with little or no experience, limited resources, and high potential for innovation to enter the market. Subjects agreed to take part in the study and filled out a characterization form mentioning their experience with Software Engineering, HCI, and User Experience Design (UX). For each of the topics, we categorized the subjects' experiences as:

\begin{itemize}
    \item None: subjects who do not know about the topic;
    \item Low: subjects with little knowledge of the topic, usually learned in courses or lectures.
    \item Medium: subjects with some knowledge of the topic and who have already carried out practical activities in coursework.
    \item High: subjects who already have some knowledge of the topic but no practical experience in the industry.
    \item Very high: subjects with advanced knowledge and practical experience with the topic in the software industry.
\end{itemize}

\textbf{Experimental design.} We used the design type \textit{one factor with two treatments}, where the factor was the wireflow creation, and the two treatments used the Startflow method and a simple paper prototyping process. We used a \textit{completely randomized design}, dividing the subjects into two groups with four subjects: the experimental group (Startflow method) and the control group (paper prototyping process). Based on the characterization form, we balanced the groups to obtain the same expertise. Table \ref{tab:controlled-exp} shows the experience of the subjects of each group.

\begin{table}[ht]
    \footnotesize
    \centering
    \begin{tabular}{ccccc}\hline
     \textbf{Group} & \textbf{Subjects} & \textbf{SE} & \textbf{HCI} & \textbf{UX}\\\hline
     \multirow{4}{6em}{Experimental} & S1 & Medium & Low & Low\\
      & S2 & Medium & Low & None\\
      & S3 & Medium & Low & Low\\
      & S4 & Medium & Low & Low\\\hline
      \multirow{4}{6em}{Control} & S5 & Medium & Low & Low\\
      & S6 & High & None & Low\\
      & S7 & Low & Medium & Medium\\
      & S8 & Low & Medium & Low\\\hline
      \multicolumn{5}{p{11cm}}{\footnotesize{Legend: \textbf{ES} - Experience in Software Engineering; \textbf{HCI} - Experience in Human-Computer Interaction; \textbf{UX} - Experience in User Experience.}}\\
    \end{tabular}
    \caption{Distribution of participants’ prior experience in subjects across the experimental and control groups.}
    \label{tab:controlled-exp}
\end{table}

\textbf{Instrumentation.} We used several instruments in this experiment: an informed consent form (ICF), a characterization form, a document containing the context of the problem, user stories (USs) and business rules (BRs) for the solution, and a guide to using the StartFlow method (for the experimental group). Additionally, we administered a questionnaire based on the Technology Acceptance Model (TAM) \cite{davis1989perceived} to gather the participants’ perceptions after completing the wireflows. The TAM questionnaire consisted of 11 questions, rated on a Likert scale from 1 (strongly disagree) to 5 (strongly agree). The questions were designed to evaluate the participants' perceived ease of use, usefulness, and overall satisfaction with the paper prototyping process, which was consistently referred to in neutral terms to avoid introducing bias, especially between the control and experimental groups. The specific questions asked were:

\begin{table}[ht]
\footnotesize
\centering
\begin{tabular}{cp{12cm}}
\toprule
\textbf{ID} & \textbf{Question} \\ 
\midrule
Q1  & Using paper prototyping improves development performance. \\
Q2  & Using paper prototyping increases productivity. \\
Q3  & Using paper prototyping improves effectiveness in development. \\
Q4  & I found the interaction with paper prototyping clear and understandable. \\
Q5  & Using paper prototyping does not require much mental effort. \\
Q6  & I find paper prototyping easy to use. \\
Q7  & I find it easy to use paper prototyping to achieve what I want. \\
Q8  & I find using paper prototyping enjoyable. \\
Q9  & The process of using paper prototyping is pleasant. \\
Q10 & I had fun using paper prototyping. \\
Q11 & Assuming I have knowledge of paper prototyping and an appropriate use case, I intend to use it again. \\
\bottomrule
\end{tabular}
\caption{Technology Acceptance Model (TAM) questionnaire administered after wireflow completion.}
\label{tab:tam_questions}
\end{table}

This questionnaire was administered to both the control and experimental groups after they completed their wireflows to gather feedback on their experience with the prototyping process.

\textbf{Preparation.}\label{controlled-preparation} We conducted a one-hour presentation to all subjects about paper prototyping and wireflows. The presentation included slides about user interface characteristics, usability and user experience, and the benefits of prototyping. In order to avoid any bias between the groups, we consistently used the term ``paper prototyping" during this presentation, even for the experimental group using the StartFlow method. The term ``paper prototyping" referred exclusively to the physical medium (paper sketches), while the activity followed the StartFlow definition of paper wireflows. This distinction was emphasized to avoid bias in the participants' interpretation of the method. For each group, we presented the specific methods (StartFlow for the experimental group and traditional paper prototyping for the control group) and instructions for application.

\subsubsection{Execution}\label{controlled-execution}

The experiment was conducted in person taking both groups of participants separately. First, the experiment was conducted with the control group. The process described in subsection \ref{controlled-preparation} was followed. The first author performed the presentation, and afterwards, the second author (moderator), has provided all the instruments and supervised the creation of the wireflows. The time each participant spent creating the wireflows was recorded for further analysis. Following the completion of this step, the TAM questionnaire was administered individually to each participant.

The same procedure was followed for the experimental group the following day. All subjects of both groups gave their feedback about the process of paper prototyping using the same TAM questionnaire to be answered by each group for consistency and comparison.

\subsubsection{Results and Discussion}

\textbf{Wireflow analysis.} The experiment provided insights into the differences in usability and task completion efficiency between the two groups. Both groups required the same number of actions to complete tasks like adding certificates or checking groups, but the experimental group, which used StartFlow, showed improvements in interface intuitiveness and user guidance.

\begin{figure}[ht]
    \centering
    \begin{minipage}[b]{0.4\linewidth}
        \centering
        \includegraphics[width=\linewidth]{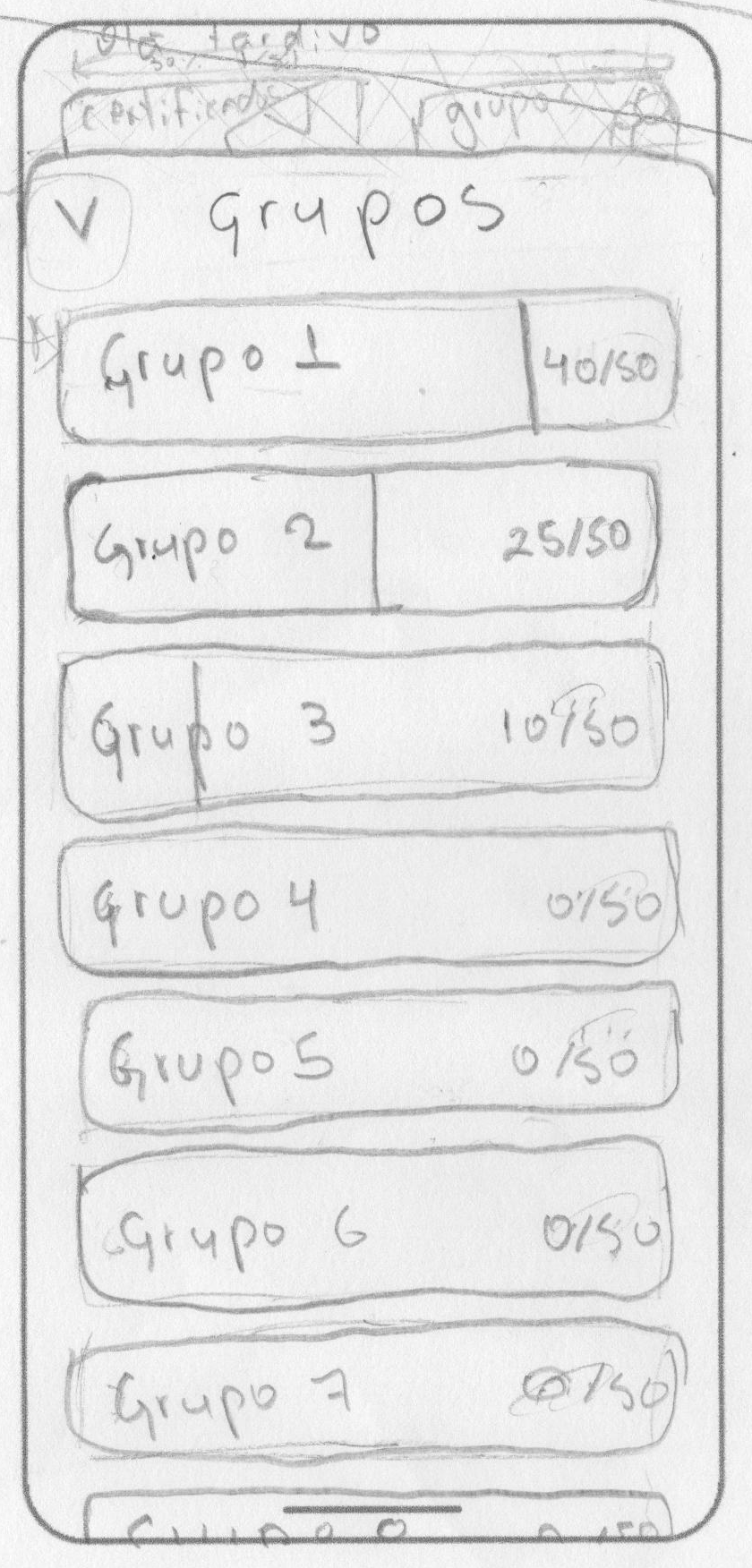}
        \caption{CAA categories screen from the control group}
        \label{fig:controle-grupos}
    \end{minipage}
    \hspace{0.05\linewidth}
    \begin{minipage}[b]{0.4\linewidth}
        \centering
        \includegraphics[width=\linewidth]{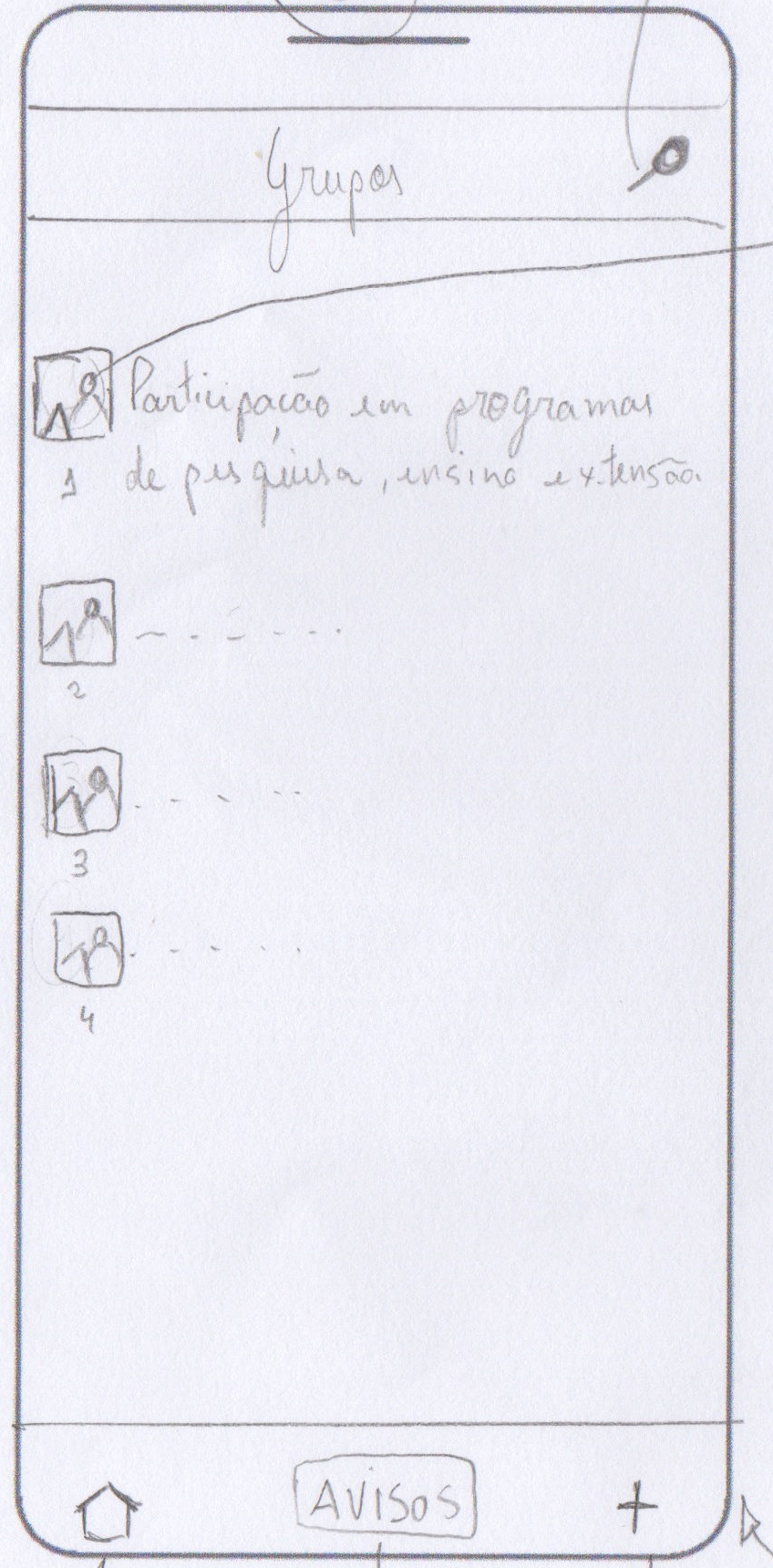}
        \caption{CAA categories screen from the experimental group}
        \label{fig:experimental-grupos}
    \end{minipage}
\end{figure}

The experimental group benefitted from the structured framework of StartFlow, which simplified task completion by offering clearer design patterns and more intuitive user flow. The control group, on the other hand, showed increased ambiguity and confusion in key aspects of the interface, particularly in relation to the management of CAA categories and the use of progress bars. An example of this can be observed when comparing the CAA categories screens of both groups, as shown in Figures~\ref{fig:controle-grupos} and~\ref{fig:experimental-grupos}. 

In the interface for the control group (Figure~\ref{fig:controle-grupos}), categories had been labeled simply as ``Group 1", ``Group 2", ``Group N", without any explanation of what the groups actually were. Without this knowledge of context, users have to memorize or assume what the specific criterion for each category is which leads to increased cognitive load and mistakes. Therefore, this can be described as a fragmented process, where the users were forced to look elsewhere or depend on pre-obtained information on what was done and what was outstanding.

By contrast, the CAA categories menu of the experimental group (Figure~\ref{fig:experimental-grupos}) was more intuitive. Each type had a title label and had more details on what exactly fit within that category. The label was also clickable for each category to enter into a more detailed view listing which of their completed activities had been assigned to said category. It also further went ahead to suggest future events, lectures, and activities that best fit the set criteria of the category, hence giving worthwhile recommendations without necessarily having the user leave the current context. This by far reduced the user's effort and reduced chances of errors during the organization of academic activities.

Overall, while the control group’s interface lacked some critical information, the experimental group, using StartFlow, appeared to offer a more informative and supportive user experience. The experimental design seemed to improve navigation clarity and provided users with actionable insights. These observations suggest that the experimental group’s wireflows may have aligned more closely with the defined user stories and business rules, though further validation in different contexts would be required to confirm these findings.

\textbf{Impact of StartFlow on Wireflow Creation.} The structured approach provided by StartFlow had an impact on wireflow creation, particularly in improving clarity and reducing the occurrence of severe usability issues. As indicated in the task comparison (Table~\ref{tab:task-completion}), both groups required the same number of steps to complete tasks. 
The experimental group’s wireflows featured clearer navigation paths, more intuitive labels for groupings, and a superior feedback system.

\begin{table}[ht]
    \footnotesize
    \centering
    \begin{tabular}{cccc}\hline
     \textbf{Task} & \multicolumn{2}{c}{\textbf{Number of Actions}}\\\hline
     \textbf{} & \textbf{Control Group} & \textbf{Experimental Group} \\ \hline
     Add Certificate & 5 & 5 \\ 
     Check Groups & 2 & 2 \\ 
     Check Certificates & 2 & 2 \\ \hline
    \end{tabular}
    \caption{Comparison of task completion actions between groups}
    \label{tab:task-completion}
\end{table}

Despite these improvements, some issues of inconsistency remained in the experimental group’s designs. For instance, while navigation was generally improved, certain interface elements, such as the placement of buttons and progress indicators, lacked uniformity across screens. This suggests that although StartFlow aided in the overall design structure, it did not entirely resolve issues related to interface consistency.

\begin{figure}[H]
    \centering
    \includegraphics[width=0.8\linewidth]{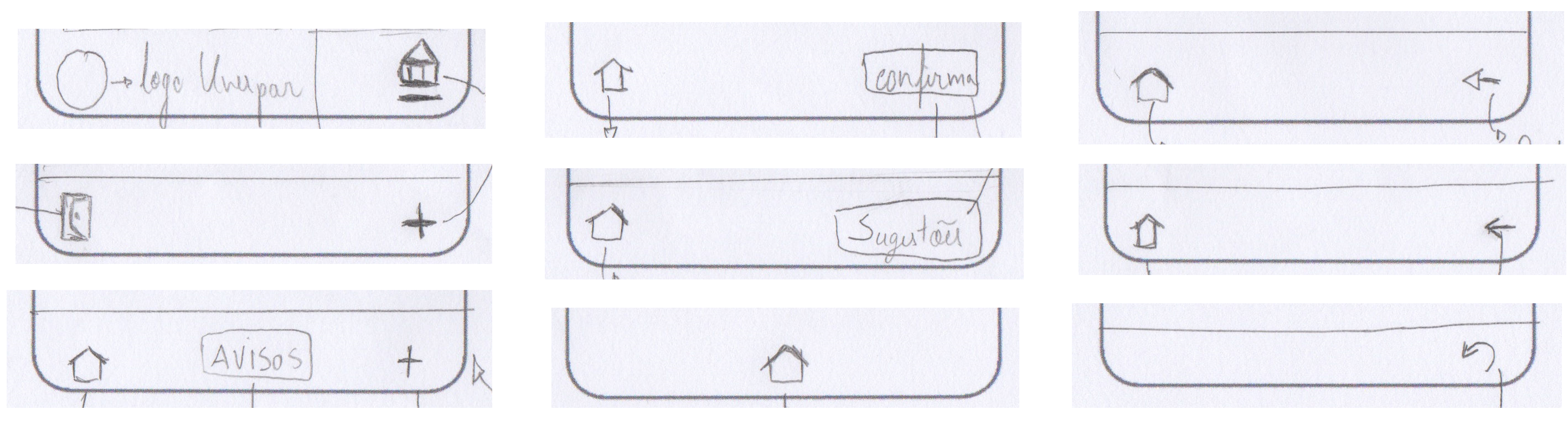}
    \caption{Inconsistent bottom navigation bar in the experimental group's wireflow}
    \label{fig:barra-inferior}
\end{figure}

A specific example of this inconsistency can be seen in the app’s bottom navigation bar, as shown in Figure~\ref{fig:barra-inferior}. Elements such as the home button and back button frequently changed positions or even disappeared across different screens. This inconsistent behavior could confuse users, as they might rely on stable navigation elements to guide their interaction with the app. Such inconsistencies indicate that while the experimental group's design was more structured overall, there were still significant areas where consistency could be improved.

\textbf{TAM Results.} The results from the TAM questionnaire added important insights into perceived usability and effectiveness, showing overall satisfaction about the whole process of prototyping in both control and experimental groups. The experimental group consistently rated their experience higher across several categories, suggesting that the structured nature of the StartFlow method positively influenced their perceptions. However, given the small sample size, the TAM responses were analyzed only descriptively. The mean scores provided in the results serve as preliminary indicators and not as conclusive evidence.


\subsubsection{Perceived Usefulness}

Perceived Usefulness (PU) is defined as “the degree to which a person believes that using a particular system would enhance their job performance” \cite{davis1989perceived}. It was measured by Q1 (“improves development performance”), Q2 (“increases productivity”), and Q3 (“improves effectiveness”). Both groups rated Q1 identically (4.25), indicating that they equally perceived prototyping as performance-enhancing. However, for Q2 the experimental group scored 4.50 versus 4.25 for the control group, suggesting a modest PU advantage when using StartFlow. Both groups again converged on 4.75 for Q3, reinforcing that the overall usefulness of the prototyping process was high across methods.

\subsubsection{Perceived Ease of Use}

Perceived Ease of Use (PEOU) is “the degree to which a person believes that using a particular system would be free of effort” \cite{davis1989perceived}. We captured PEOU via Q4 (clarity of interaction), Q5 (mental effort), Q6 (ease of use), and Q7 (goal achievement). For Q4, the experimental group rated 4.75 compared to the control group’s 4.50, indicating clearer interaction afforded by StartFlow’s guidance. On Q5 (mental effort), the experimental group’s lower score (3.00 vs. 3.25) suggests that the structured framework slightly reduced cognitive load. Q6 was equal (4.50), reflecting similar basic usability, while Q7 dipped for the experimental group (4.00 vs. 4.50), implying that the method’s structure sometimes felt restrictive rather than liberating.

\subsubsection{Perceived Enjoyment}

Perceived Enjoyment (PE) refers to “the extent to which the activity of using a specific system is perceived to be enjoyable in its own right, aside from any performance consequences” \cite{van1992intrinsic}. This intrinsic motivation dimension was measured by Q8 (“enjoyable”), Q9 (“pleasant”), and Q10 (“I had fun”). The experimental group scored identical to control on Q8 and Q9, but saw a notable rise on Q10 (4.75 vs. 4.00), indicating that StartFlow’s structure fostered greater engagement and fun, consistent with findings that structured guidance can enhance intrinsic enjoyment in novice users \cite{chesney2006perceived}.

\subsubsection{Behavioral Intention to Use}

Behavioral Intention (BI) in TAM is “the degree to which a person has formulated conscious plans to perform or not perform some specified future behavior” \cite{venkatesh2000model}. Q11 (“intend to use again”) directly taps this construct. The experimental group’s BI score of 4.75 versus 4.00 for the control group demonstrates that participants who used StartFlow are more likely to adopt the method in the future, reflecting how enhanced PU, PEOU, and PE jointly elevate intention to use.

\subsection{Heuristic Evaluation}\label{heuristic-evaluation}

Following the creation of the artifacts in the experiment, the second phase of our proof of concept was a heuristic evaluation of the generated wireflows. The goal of this phase, framed using the GQM paradigm, was to: ``Analyze the wireflows produced by the control and experimental groups for the purpose of evaluation with respect to their usability quality from the point of view of UX experts in the context of a heuristic evaluation.''

\subsubsection{Planning}\label{heuristic-planning}

The planning of the heuristic evaluation focused on establishing a process to assess the quality of the wireflows. This involved defining the objective of the evaluation, variables, evaluator selection criteria, the evaluation protocol, and the instruments to be used.

\textbf{Objective.} The primary objective was to quantitatively compare the wireflows produced by the control and experimental groups. The comparison focused on identifying and classifying usability defects to determine whether the use of the StartFlow method resulted in higher quality artifacts.

\textbf{Variables.} The independent variable for this evaluation was the \textbf{group}, either control or experimental, that produced the wireflow. The dependent variables were the quantitative measures of usability quality collected during the evaluation such as the total number of identified usability defects, their severity, the specific Nielsen Heuristic violated by them and their location within the prototype application's interface.

\textbf{Selection of Evaluators.} We invited nine specialists to act as evaluators. All participants held a Ph.D. in Software Engineering or related fields and possessed research and practical experience in usability and UX. This level of expertise was important for ensuring a reliable evaluation.

\textbf{Evaluation Protocol.} The evaluation was carried out using a structured protocol. All wireflows were anonymized so that evaluators would be blind to which group produced which artifact, thus mitigating potential bias. The evaluation itself was based on Jakob Nielsen's 10 Usability Heuristics \cite{nielsen1994usability}, a widely accepted standard. The evaluators were instructed to follow Nielsen's heuristics to identify defects, provide a written justification for each, and rate their severity using a 4-point scale, as detailed in Table \ref{tab:severity_scale}. Each defect should be associated with one or more heuristics that allowed it to be identified.

\begin{table}[ht]
\footnotesize
\centering
\begin{tabular}{ccp{9cm}}
\toprule
\textbf{Rating} & \textbf{Level} & \textbf{Description} \\ 
\midrule
1 & Cosmetic & Low-priority defect that does not obstruct functionality. \\
2 & Minor & Small defect with moderate impact on the user experience. \\
3 & Major & Serious defect whose correction is a high priority. \\
4 & Catastrophic & Critical defect that must be fixed for the interface to be usable. \\
\bottomrule
\end{tabular}
\caption{Defect Severity Scale Used by Evaluators}
\label{tab:severity_scale}
\end{table}

\textbf{Instrumentation.} The evaluators were equipped with a set of materials to guide their analysis, including the anonymized wireflows from both groups, a document detailing the user stories and business rules (Table \ref{tab:us_br}) provided to the students, a structured evaluation form to record defects, justifications, heuristic violations, and severity ratings. After evaluation, the data was compiled into spreadsheets for quantitative analysis.

\subsubsection{Execution}\label{heuristic-execution}

The execution of the heuristic evaluation was carried out remotely. They were instructed to independently review the two sets of anonymized wireflows. For each wireflow, they performed a thorough analysis, documenting any usability defect found on the provided evaluation form. After all the evaluations were completed, the researchers collected the forms. The data was then systematically compiled into a single dataset, where defects were categorized, duplicates were identified, and the quantitative metrics such as defect counts and severity scores were calculated for each group to enable a direct comparison.

\subsubsection{Results and Discussion}\label{heuristic-results}

The analysis of the expert evaluations provided quantitative insights into the effectiveness of the StartFlow method in reducing usability defects.

\textbf{Defect and False Positive Analysis.} The total number of discrepancies identified by the experts was lower for the experimental group. Furthermore, the experimental group's wireflows generated fewer false positives, which are issues flagged by experts that are not actual defects. As shown in Table     \ref{tab:testeeee}, the experimental group had 10\% fewer real defects than the control group.

\begin{table}[ht]
    \footnotesize
    \centering
    \begin{tabular}{cccc}\hline
     \textbf{Group} & \textbf{Total Discrepancies} & \textbf{False Positives} & \textbf{Real Defects} \\ \hline
     Control & 51 & 9 & 42 \\ 
     Experimental & 42 & 4 & 38 \\ \hline
    \end{tabular}
    \caption{Summary of Discrepancies and Defects Identified by Evaluators}
    \label{tab:testeeee}
\end{table}

When distinguishing between unique defects and duplicates which can be described as the same defect found by multiple evaluators, the experimental group showed a notable reduction in unique defects, as detailed in Table \ref{tab:defects_distinction}. This suggests that StartFlow may help teams avoid introducing novel or original usability problems, although both groups produced an identical number of duplicate defects.

\begin{table}[ht]
    \footnotesize
    \centering
    \begin{tabular}{ccc}\hline
     \textbf{Group} & \textbf{Unique Defects} & \textbf{Duplicates} \\ \hline
     Control & 18 & 24 \\ 
     Experimental & 14 & 24 \\ \hline
    \end{tabular}
    \caption{Distinction Between Unique and Duplicate Defects}
    \label{tab:defects_distinction}
\end{table}

\textbf{Defect Severity.} An interesting finding emerged from the severity analysis. Although the experimental group produced fewer defects, their average severity was slightly higher (2.68) compared to the control group (2.52). This indicates that while StartFlow helped reduce the overall quantity of issues, the problems that did appear were considered more critical by the experts. This may suggest that the method helps resolve minor issues, leaving more complex, severe problems to be addressed.

\textbf{Heuristic Violations.} The distribution of defects across Nielsen's heuristics reveals how the StartFlow method influenced the types of usability problems. As shown in Table \ref{tab:heuristic_violations}, the control group's main issues were related to ``User control and freedom'' and ``Match between system and the real world.'' In contrast, the experimental group's most significant problem was ``Consistency and standardization,'' with 13 occurrences. This finding is critical, as it suggests that while StartFlow provides a helpful structure, it may not sufficiently guide non-experts in maintaining a consistent design across all interface screens.

\begin{table}[ht]
    \footnotesize
    \centering
    \begin{tabular}{lcc}\hline
     \textbf{Heuristic} & \textbf{Control Group} & \textbf{Experimental Group} \\ \hline
     Consistency and standardization & 4 & \textbf{13} \\
     User control and freedom & \textbf{10} & 3 \\
     Match between system and real world & \textbf{9} & 7 \\
     Visibility of system status & 7 & 6 \\
     Aesthetic and minimalist design & 7 & 1 \\
     Error prevention & 7 & 6 \\ \hline
    \end{tabular}
    \caption{Comparison of the Top 5 Defects per Heuristic for Each Group}
    \label{tab:heuristic_violations}
\end{table}

\textbf{Defect Location in the Prototypes' Interfaces.} The location of defects also differed between the groups, as seen in Table \ref{tab:defect_location}. The control group's problems were concentrated in the ``Main Menu'' (16 defects), a central navigation area. The experimental group, however, struggled most with the ``CAA Groups'' screen (10 defects) and the ``CAA Registration'' screen (13 defects). The ``CAA Groups'' screen was particularly complex, as it required displaying progress across multiple categories with specific rules (BR3 and BR4). The high number of defects here for the experimental group reinforces the finding that StartFlow helped with overall structure but did not fully resolve challenges in designing complex, rule heavy interface components.

\begin{table}[ht]
    \footnotesize
    \centering
    \begin{tabular}{lcc}\hline
     \textbf{Location} & \textbf{Control Group} & \textbf{Experimental Group} \\ \hline
     Main Menu & \textbf{16} & 5 \\
     CAA Registration & 13 & 13 \\
     Login & 8 & 4 \\
     Tutorial & 7 & - \\
     CAA Groups & 3 & \textbf{10} \\
     Current CAAs & 3 & 5 \\
     General & 1 & 8 \\ \hline
    \end{tabular}
    \caption{Location of Defects Found in the Prototypes' Interfaces}
    \label{tab:defect_location}
\end{table}

This quantitative data aligns with the qualitative observations. For instance, the experimental group’s difficulty with the ``CAA Groups'' screen (Figure~\ref{fig:experimental-grupos}) can be attributed to the challenge of maintaining consistency and clarity while adhering to complex business rules.

\section{Discussion}

This section presents discussions of the main findings from the focus group and proof of concept.

\subsection{Focus Group}

The focus group study findings revealed mixed perceptions regarding the applicability of the StartFlow method in the context of software startups. On the one hand, the positive aspects highlighted by participants demonstrate the method's potential as a support tool in the initial phase of interface design, encouraging reflection on functionality, requirements, and user flow. This result complements studies indicating that lightweight, structured methods or tools can serve as cognitive support for novice designers \cite{schneider2016crowdux} and help organize design reasoning when UX specialists are not available \cite{ovad2016reduce, ovad2016templates}. The use of questioning points, the flexibility in using artifacts and tools, and the clarity of the process steps were highlighted as important differentiators, favoring its adoption in academic settings or by teams that already have some maturity in defining functionality. These points indicate that the method can help initiate communication between designers and developers and may allow for limited reuse of visual components in different projects.

On the other hand, the complaints raised, especially by the ``haters" group, reveal important limitations that need to be considered for StartFlow to achieve greater traction in the dynamic and uncertain environment of startups. The rigidity and linearity of the method contrast with the iterative and experimental practices standard in this type of organization. This divergence highlights the need to make the possibility of iterative cycles and incremental method use more explicit, which could be addressed through improvements in documentation and visual representation of the steps. This finding confirms recurring criticisms of prescriptive approaches in agile contexts, where overly sequential methods are often perceived as misaligned with startups' organizational cultures \cite{ghezzi2018formal,lermen2023does}. At the same time, the result broadens this discussion by showing that resistance may not be associated solely with the method's structure but also with its presentation. Criticism of the predominance of textual content highlights a usability barrier, particularly in multidisciplinary teams that value more intuitive visual representations. Furthermore, the lack of clear guidelines for evaluating the method's effectiveness and the weak connection to the user experience suggest that StartFlow, in its current form, may be perceived more as a tool to support interface design than as a user-centered approach in a broader sense. This result aligns with an interesting line of research that argues that design methods should also be considered and evaluated from the perspectives of usability and user experience \cite{nakamura2021ux, rajeshkumar2013taxonomies}, especially when aimed at non-specialist professionals \cite{ovad2016templates}.

Another relevant point concerns the vocabulary used. The use of the term "wireflow," for example, was perceived as potentially too technical for teams without UX training, which could hinder their understanding and adoption of the method in broader contexts. This finding reinforces studies on the democratization of design, which point to specialized language as an obstacle to the dissemination of UX practices across multidisciplinary teams \cite{ibargoyen2013elephant}. Thus, terminological adaptation and the use of accessible examples are critical factors in expanding the method's adoption.

Thus, the discussion of the results highlights a duality: although StartFlow is recognized for its practical value and ease of use, its widespread adoption depends on significant adjustments to its presentation, flexibility, and alignment with iterative practices, evaluation metrics, and broader UX principles. 

\subsection{Proof of Concept}

The proof of concept conducted in this study aimed to verify whether the StartFlow method supports professionals without UX expertise, especially in the context of early-stage software startups. The results of the experiment and the heuristic evaluation by experts provide preliminary evidence that the method can positively affect the quality of the artifacts produced and the user experience associated with the prototyping process.

Initially, the analysis of the wireflows generated by the teams showed that the experimental group (using StartFlow) demonstrated greater adherence to the user stories and defined business rules. The clarity of the information, the presence of appropriate feedback, the organization of categories, and the more intuitive user guidance indicate that the method contributed to a more accurate and usable representation of the MVP. For example, while the control group presented generic and confusing nomenclature ("Group 1", "Group 2," etc.), the experimental group was able to represent categories with informative, interactive, and contextualized labels, in addition to including proactive suggestions for user actions. This result aligns with previous studies that highlight the importance of system state visibility and real-world consistency as principles of a good UX \cite{guerino2024wireflows}.

In addition to the qualitative analysis of the prototypes, the data collected via the TAM questionnaire supports this trend. Participants who used StartFlow gave higher scores in categories such as perceived clarity (Q4), enjoyment (Q10), and intended future use (Q11). The structure provided by the method appears to have generated a sense of support, guidance, and purpose in participants, which aligns with research indicating that structured processes with heuristic checkpoints promote engagement and confidence among novice UX professionals \cite{hokkanen2015ux, hokkanen2016minimum}.

On the other hand, we highlight that, although the experimental group had a lower total number of defects and false positives in the heuristic evaluation, the average severity of the defects found was slightly higher (2.68 versus 2.52). This finding suggests that the method helped eliminate superficial problems, but did not prevent the occurrence of more critical errors, possibly related to the complexity of business rules and the participants' inexperience with visual consistency and standardization of interface elements. This result is consistent with research indicating that low-fidelity prototyping can limit the capture of problems related to visual consistency and element standardization, especially when conducted by inexperienced participants \cite{da2025new,zhang2022speeding}. This finding was evident, for example, in the inconsistency of the bottom navigation menu, whose absence or displacement on different screens created potential points of confusion for users. 

Furthermore, the heuristic evaluation data showed that the StartFlow method can directly influence the type of errors committed. While the control group had more problems related to user freedom and correspondence with the real world, the experimental group focused on inconsistency and standardization. This finding shows that StartFlow, by supporting the precise definition of functionalities and flows, helps prevent errors related to the system's usage logic; however, it still lacks mechanisms to ensure visual and aesthetic consistency, a particularly sensitive dimension in contexts with multiple participants and without previously established standards.

Finally, the results show that StartFlow's main contribution lies not in reducing the absolute number of steps or clicks required to complete tasks, but in the quality of the flow and the way interactions are planned and organized. The equivalence in the number of actions between the groups (Table 6) indicates that the method does not necessarily "shorten" the path, but rather makes it more understandable and safer for the user. This is consistent with the method's objectives, which aim to structure design thinking in a user-centered and heuristic-driven manner, even in low-fidelity contexts and with scarce resources. This result reinforces studies arguing that UX efficiency should not be measured solely by quantitative metrics, such as the number of clicks, but also by predictability, comprehensibility, and perceived safety for users \cite{deters2024x,obaidi2025app}.


In summary, our initial findings indicate that StartFlow can help novice participants structure clearer navigation flows and better align prototypes with defined requirements. Despite the persistence of critical defects and some consistency issues, the results provide early evidence that the method may offer value for guiding low-fidelity prototyping. Further studies with startup practitioners are needed to determine its applicability in professional settings.

\subsection{Implications}

The findings of this two studies point to implications for UX practice, research, and teaching in technological innovation contexts. First, StartFlow emerges as a viable tool for startup environments with lean teams unfamiliar with user-centered design but under significant pressure to deliver value quickly. Its step-by-step structure, with guiding questions and heuristics, represents a practical resource for facilitating design decisions at low cost and without relying on experts.

In academia, the results indicate that methods like StartFlow can be explored in introductory UX or software engineering courses, offering a teaching experience that combines structure, creativity, and continuous validation. For research purposes, this study paves the way for future investigations into the integration of StartFlow with other agile approaches and visual prototyping techniques, as well as its application in more diverse domains and with multidisciplinary teams.

Finally, there are implications for the development of tools to support the method. The StartFlow framework is compatible with digital prototyping platforms, and its use could be enhanced with templates, interactive wizards, or plugins that visually guide the application of heuristics.

\subsection{Threats to Validity and Limitations}

Regarding the focus group, one of the threats identified is the generalization of the results \cite{deFranca2015}. However, this threat is inherent to qualitative studies conducted using this method, as the responses are closely linked to the subjectivity of the participants \cite{deFranca2015}. To mitigate this threat, we conducted the study with a recommended number of people for focus groups \cite{Kontio2008}. Another threat is related to the participants' learning about StartFlow. The method was explained and made available to the participants only during the study. Therefore, it is possible that we would have obtained different results if the documentation had been sent to the participants in advance and a preliminary study had been conducted before the study began. Regarding data analysis, to obtain data on usefulness, ease of use, and comprehension, we attempted to triangulate the comments made by the groups with the \textit{post-its} that were inserted in the boards. In this way, every comment they made was based on an opinion written on the \textit{post-it} that was discussed as a group.

About the proof of concept, despite the promising results, this research has some limitations that should be considered. First, the number of participants in the experiment (eight students) is small, which limits the generalizability of the findings. Although the profile of computer science students resembles that of members of early-stage startups, the lack of diversity in backgrounds and contexts may bias the results. Moreover, although we explained during preparation that paper prototyping referred only to the physical medium, the group that used the Startflow method may also have had an interpretative bias regarding the concept. not account for relevant external factors such as real-world deadlines, client pressures, or market uncertainties. Furthermore, the proposed task was somewhat simple and focused on a specific domain (academic activity management), which may have facilitated the application of the method.

Another relevant point is that StartFlow does not include direct mechanisms to ensure visual standardization between screens, which was evident in the results of the heuristic evaluation. Therefore, although it facilitates the creation of coherent flows, the method could benefit from improvements aimed at visual cohesion of prototypes. Additionally, because the data were collected from self-assessments (in the case of TAM) and subjective inspections (in the subject of heuristic evaluation), there is a risk of bias in interpretations. Future studies could explore additional objective measures, such as task completion times by real users or analysis of usability metrics in controlled tests.

\section{Conclusion and future work}

This article introduces the StartFlow method, a structured approach designed to support software startup professionals in building MVP prototypes using the wireflow technique. Through several preliminary studies, we developed the method with a focus on the reality of early-stage startups, which often lack UX specialists, financial resources, and well-defined processes. By proposing a sequence of straightforward steps, accompanied by guiding questions and heuristic criteria inspired by Nielsen, StartFlow seeks to promote the practical application of usability and user experience principles in an accessible, streamlined, and iterative manner.

Before conducting a proof of concept, we held a focus group with researchers in Software Engineering, HCI, and Software Startups to gather initial feedback and assess the method's perceived usefulness, ease of use, and comprehensibility. Participants highlighted several strengths of StartFlow, including its support for identifying product features through question prompts, its contribution to organizing user flows and visual components, and its flexibility to accommodate different types of requirements and prototyping tools. However, relevant criticisms were also raised, particularly regarding the method's presentation and perceived sequentiality, which could conflict with the agile and dynamic nature of startups. Some participants suggested making its iterative potential more explicit by incorporating more visual materials into the documentation and strengthening its alignment with broader UX concerns beyond interface design.

The proof-of-concept, which consisted of an experiment with Computer Science students and a heuristic evaluation with experts, demonstrated that the use of StartFlow resulted in more precise prototypes, aligned with business rules, and better evaluated for usability. Furthermore, the method was well received by participants, indicating ease of use, clarity, and intended future use. Although the method did not eliminate all visual consistency issues, we observed a reduction in minor defects and an improved structuring of the navigation flow, demonstrating its potential in real-world startup contexts.

As future work, we are incorporating the improvements identified from the focus group and proof of concept results into a new version of the method. Moreover, our work opens doors for possible future research: (i) application of StartFlow in real startups: replicating the study with professionals working in real startups can provide more robust evidence on the method's effectiveness and adaptability in high-pressure environments, with real clients and market constraints; (ii) integration with digital prototyping tools: developing supporting artifacts, such as templates, interactive guides, or plugins for tools like Figma and Balsamiq, can further facilitate the method's adoption, including by distributed teams; (iii) extension of the method with a focus on visual consistency: considering that the heuristic evaluation indicated standardization issues in the generated prototypes, it is pertinent to investigate how StartFlow can be improved with additional guidance on style, visual alignment, and reusable components; (iv) longitudinal and collaborative studies: investigating how the method behaves over time, in iterative design and development cycles, as well as its applicability in multidisciplinary teams with diverse profiles, can generate important insights for its evolution; (v) analysis of the method's impact on UX learning: given its acceptance among participants without prior UX training, StartFlow can be explored as a pedagogical resource in introductory courses in interaction design or software engineering, which also deserves further investigation. 

With these perspectives, we hope that StartFlow will contribute to both professional practice in startups and the advancement of knowledge about methods to support prototyping and user experience in technological innovation contexts.

\bibliographystyle{elsarticle-num} 
\bibliography{references}

\end{document}